\documentclass[aps,prl,twocolumn,preprintnumbers,showpacs,superscriptaddress,nofootinbib,floatfix,10pt]{revtex4-1}
\pdfoutput=1
\usepackage{textcomp}
\usepackage[english]{babel}
\usepackage{amsmath,amssymb,amsbsy,booktabs}
\usepackage{bm}
\usepackage{color}
\usepackage{array}
\usepackage{amstext}
\usepackage{graphicx}
\usepackage{amsfonts}
\usepackage{bm}
\usepackage{dcolumn}
\usepackage{rotating}
\usepackage{epstopdf}
\usepackage{esint}
\usepackage{url}
\usepackage{slashed}
\usepackage[colorlinks=true,
linkcolor=blue,
filecolor=blue,
anchorcolor=blue,
urlcolor=blue,
citecolor=blue
]{hyperref}

\usepackage[T1]{fontenc} 
\usepackage[utf8]{inputenc} 
\usepackage{float}
\usepackage{subfigure}

\allowdisplaybreaks[0]

\begin{document}
	
	\title{\bf Baryogenesis from Hierarchical Dirac Neutrinos}
	\author{Shao-Ping Li}
	\email{ShowpingLee@mails.ccnu.edu.cn}
	\affiliation{Institute of Particle Physics and Key Laboratory of Quark and Lepton Physics~(MOE),\\
		Central China Normal University, Wuhan, Hubei 430079, P.~R.~China}
	
	\author{Xin-Qiang Li}
	\email{xqli@mail.ccnu.edu.cn}
	\affiliation{Institute of Particle Physics and Key Laboratory of Quark and Lepton Physics~(MOE),\\
		Central China Normal University, Wuhan, Hubei 430079, P.~R.~China}
	
	\author{Xin-Shuai Yan}
	\email{xinshuai@mail.ccnu.edu.cn}
	\affiliation{Institute of Particle Physics and Key Laboratory of Quark and Lepton Physics~(MOE),\\
		Central China Normal University, Wuhan, Hubei 430079, P.~R.~China}
	
	\author{Ya-Dong Yang}
	\email{yangyd@mail.ccnu.edu.cn}
	\affiliation{Institute of Particle Physics and Key Laboratory of Quark and Lepton Physics~(MOE),\\
	    Central China Normal University, Wuhan, Hubei 430079, P.~R.~China}
	
\begin{abstract}
We present here a novel and testable mechanism for leptogenesis, which is characterized by a purely thermal generation of lepton asymmetry via scalar decay. Guided by the thermal mass matrix diagonalization in the finite-temperature regime, we propose a scenario in which the baryon asymmetry is formulated in terms of the masses and mixing from hierarchical Dirac neutrinos, as well as a vacuum scale accountable for the sub-eV neutrino masses. This allows a natural and direct link between the high-scale \textit{CP} asymmetry and the low-energy Dirac \textit{CP}-violating phase without involved model structures, and thus circumvents the haunting problem encountered in general leptogenesis scenarios. The mechanism can also be applied as a prototype for solving the baryon asymmetry problem to a broad class of well-motivated model buildings.
\end{abstract}
	
	
	\maketitle
	
\textit{Introduction.}--The baryon asymmetry in the Universe (BAU) represents one of the biggest conundrums brought to the Standard Model (SM) of particle physics. As the mainstream to explain the BAU, the leptogenesis~\cite{Fukugita:1986hr} exhibits however a haunting problem in establishing a direct link to the low-energy leptonic \textit{CP}  violation~\cite{Buchmuller:1996pa,Branco:2001pq,Branco:2002kt}. Thus far, an unambiguous connection is still missing because not all the participated Yukawa couplings can be uniquely fixed by the observed lepton masses and the Pontecorvo-Maki-Nakagawa-Sakata (PMNS) matrix. 

This well-known obstacle has triggered many attempts in the popular seesaw-based leptogenesis~\cite{Fukugita:1986hr,Akhmedov:1998qx,Pilaftsis:2003gt,Hambye:2016sby} to source the baryon asymmetry uniquely from the \textit{CP}-violating phase in the PMNS matrix (we call this connection the \textit{unique phase source} hereafter), where assumptions and/or parametrizations of the neutrino Dirac-Yukawa structure are generically made, such as the widely used Casas-Ibarra parametrization~\cite{Casas:2001sr}; see also some earlier and recent considerations~\cite{Buchmuller:2000as,Joshipura:2001ui,Davidson:2001zk,Moffat:2018smo,Xing:2020ghj,Xing:2020erm,Rahat:2020mio}. Concerning the unique phase source problem, similar situations are also expected in the Dirac leptogenesis~\cite{Dick:1999je}, where the desired link can only be realized by making assumptions and/or parametrizations of unknown Yukawa structures~\cite{Gu:2006dc,Gu:2007mi,Wang:2016lve,Li:2020ner}. A more comprehensive review in this context can also be found in Ref.~\cite{Branco:2011zb}. Such an obstacle has prompted a notorious conclusion: leptogenesis cannot \textit{unambiguously} declare a successful BAU resolution in light of the leptonic \textit{CP} violation to be confirmed at low-energy experiments~\cite{Branco:2001pq,Rebelo:2002wj,Branco:2011zb}.

Without further high-energy assumptions or explicit Yukawa structures, it is quite challenging to realize the unique phase source in leptogenesis scenarios. Despite the known obstacles, the significance to establish a direct link between the BAU mystery and the Dirac \textit{CP}-violating phase in the PMNS matrix is dramatic and can at least be manifested in a threefold way. First of all, such a direct connection indicates that, without resorting to the unknown \textit{ultraviolet} phases and sophisticated model structures, the to-be-determined Dirac \textit{CP} phase in neutrino oscillation experiments can entirely source the BAU. Given that a \textit{CP}-violating signal in the leptonic sector is hinted by the recent T2K observation~\cite{Abe:2019vii}, a successful link is now becoming even more appealing. In addition, the simple formulation can be readily tested in light of neutrino oscillation experiments. Finally, the link to Dirac rather than Majorana \textit{CP}-violating phase may also help us to infer the neutrino properties.
 
In this Letter, we  will apply the thermal field theory~\cite{Das1997} to the idea of Dirac leptogenesis, and present for the first time a scenario to realize the unique phase source without bothering high-energy flavor-model buildings on the unknown   Yukawa structures. Under the purely thermal Dirac leptogenesis, we will show in the scenario that it is able to formulate the baryon asymmetry generation in terms of the detectable neutrino oscillation observables, as well as a restricted vacuum scale that can partially explain the smallness of Dirac neutrino masses via a seesaw-like relation. Therefore, the  scenario  can closely connect the Dirac nature of neutrinos with the BAU problem in an experimentally testable way.

It should be pointed out that the accumulation of lepton asymmetry and the sphaleron reprocessing to baryon asymmetry share the basic idea of the canonical Dirac leptogenesis~\cite{Dick:1999je}. Explicitly, the \textit{CP} asymmetry is generated by an out-of-equilibrium scalar decay, where feeble Dirac neutrino Yukawa couplings, being smaller than the charged lepton ones, ensure a late left-right equilibration (LRE) after sphaleron freezes out, and the final baryon asymmetry is free from significant flavor effects, since it is predominantly determined by the asymmetry stored in the right-handed neutrino species at the sphaleron decoupling regime. However, the  mechanism presented here is based on \textit{purely thermal effects}~\cite{Hambye:2016sby,Giudice:2003jh}. Particularly, the one-loop self-energy diagram can produce kinetic phase with the thermal cutting rules, which would otherwise vanish in the zero-temperature quantum field theory with the usual on-shell cut. It is the purely thermal effects in the Dirac leptogenesis that make us free from unknown Yukawa couplings, specific flavor structures, and/or new degrees of freedom at the scale of possible grand unified theories, and finally lead us to realize the unique phase source in a natural way.

\textit{Diagonal thermal mass basis.}--Let us illustrate the underlying ingredients leading to the direct BAU-PMNS connection from a minimal Lagrangian density:
\begin{align}\label{lag}
	-\mathcal{L}=\tilde{Y}_\ell \bar L \phi e_R+ \tilde{Y}_\nu \bar L \tilde{\Phi}\nu_R+\mathrm{H.c.},
\end{align} 
where $\phi$ is the SM-like Higgs doublet, and $\tilde{\Phi}\equiv i\sigma_2\Phi^*$ with $\Phi$ being a new Higgs doublet. The Dirac neutrinos become massive after $\Phi$ develops a smaller vacuum expectation value, $\langle \Phi \rangle =(0,v_\Phi/\sqrt{2})^T$. Note that, to generate the sub-eV Dirac neutrino masses while satisfying at the same time the proper LRE condition $\tilde{Y}_\nu\lesssim\mathcal{O}(m_e/v_\phi)$~\cite{Dick:1999je}, we expect  $v_\Phi\sim \mathcal{O}(\text{keV})  \ll v_{\phi}\simeq 246$~GeV. Such a vacuum hierarchy can be nicely explained if the Higgs potential exhibits a soft symmetry breaking term, $\mu^2 \phi^\dagger \Phi+\rm H.c$. With the soft-breaking dimensional parameter $\mu$ being smaller than $v_{\phi}$, a seesaw-like relation $v_\Phi v_{\phi} \simeq \mu^2$ can be induced for an electroweak-scale $\Phi$, or $v_\Phi\simeq \mu^2 v_\phi/M_\Phi^2\ll v_\phi$ for a much heavier $\Phi$~\cite{Davidson:2009ha}. Explicitly, the scalar potential satisfying these requirements can be constructed with a global $U(1)$ symmetry in the $\nu_R$-$\Phi$ sector, and is given by~\cite{Davidson:2009ha}
\begin{align}\label{pot}
	V(\phi,\Phi)&=m_{11}^2 \phi^\dagger \phi+m_{22}^2 \Phi^\dagger \Phi -\left[\mu^2 \phi^\dagger \Phi+ \text{H.c.}\right]
	\nonumber \\[0.2cm]
	&+\frac{\lambda_1}{2}(\phi^\dagger \phi^2)+\frac{\lambda_2}{2}(\Phi^\dagger \Phi)^2
		\nonumber \\[0.2cm]
	&+\lambda_3(\phi^\dagger \phi)(\Phi^\dagger \Phi)
	+\lambda_4 (\phi^\dagger \Phi)(\Phi^\dagger \phi),
\end{align}
which is softly broken by the $\mu^2$ term. The parameter $\mu^2$ can also be made real by rephasing $\Phi$, such that the CP phase necessary for the BAU problem comes exclusively from the lepton Yukawa sector. Besides, Eq.~\eqref{pot} induces a degenerate mass in the neutral component of $\Phi$,
\begin{align} \label{eq:massspec}
	m_{H,A}^2=m_{H^+}^2+\frac{\lambda_4}{2}v_{\phi}^2,
\end{align}
while the mass of charged Higgs, $m_{H^+}^2= m_{22}^2+\lambda_3 v_{\phi}^2/2$,  can also be made comparable to that of the neutral component if the parameter $\lambda_4$ is small. Such a mass degenerate pattern, together with the absence of CP violation in the scalar sector, will remove possible strongly first-order phase transition and hence the electroweak baryogenesis~\cite{Morrissey:2012db}, ensuring our purely thermal Dirac leptogenesis the very mechanism for the BAU problem.

In the finite-temperature domain where the lepton doublet ($L$) participates in the generation of leptonic \textit{CP} asymmetry, the thermal mass $m^2_{L}(T)$ can be formally written as~\cite{Weldon:1982bn}
\begin{align} 
	m^2_{L}(T)=\mathcal{I}_1(g)+\mathcal{I}_2(\tilde{Y}_\ell)+\mathcal{I}_3(\tilde{Y}_\nu),
\end{align}
where $\mathcal{I}_1(g)=\left(3g_2^2/32+g_1^2/32\right) T^2 $, with $g_2$~($g_1$) being the $SU(2)_L$~($U(1)_Y$) gauge coupling, comes from the gauge interactions and is diagonal,  
while in the flavor basis of Eq.~\eqref{lag}, the corrections $\mathcal{I}_2(\tilde{Y}_\ell)$ and $\mathcal{I}_3(\tilde{Y}_\nu)$, which arise respectively from the right-handed charged leptons $e_R$ and neutrinos $\nu_R$ running in the $L$ self-energy loop, are generically non-diagonal matrices. For thermally distributed $e_R$ and $\nu_R$ with a common temperature, we have 
\begin{align}
\mathcal{I}^{eq}_{2}= \frac{\tilde{Y}_{\ell}\tilde{Y}^\dagger_{\ell }}{16}T^2,\qquad  \mathcal{I}^{eq}_{3}= \frac{\tilde{Y}_{\nu}\tilde{Y}^\dagger_{\nu}}{16}T^2.
\end{align}
In realistic situation, however, the right-handed Dirac neutrinos accumulate only slowly their abundances via Eq.~\eqref{lag} due to the feeble $\tilde{Y}_\nu$ under the LRE condition. This condition further induces a fact that, unlike the three $e_R$ that keep thermalized via gauge interactions, $\nu_R$ would carry a much suppressed phase-space distribution $f_{\nu_R}(p)\ll f_{\nu_R}^{eq}(p)$ in the early Universe. These consequences drive a doubly suppressed contribution from $\mathcal{I}_3(\tilde{Y}_\nu)$. As a result, $m^2_{L}(T)$ is essentially given by
\begin{align}\label{ther-fer-m-2}
	\frac{m^2_{L,ij}(T)}{T^2}=\left(\frac{3}{32}g_2^2+\frac{1}{32}g_1^2\right) \delta_{ij}+\frac{1}{16}\left(\tilde{Y}_{\ell}\tilde{Y}^\dagger_{\ell}\right)_{ij},
\end{align}
while the thermal masses of $e_R$ and $\nu_R$, both of which receive finite-temperature corrections from lepton and Higgs doublets via Eq.~\eqref{lag}, are given by~\cite{Weldon:1982bn} 
	\begin{align}\label{ther-rfer-m-2}
		\frac{m^2_{e_R,ij}(T)}{T^2}&=\frac{1}{8}g_1^2 \delta_{ij}+\frac{1}{8}\left(\tilde{Y}^\dagger_{\ell}\tilde{Y}_{\ell}\right)_{ij}, \\ \label{ther-rfer-m-22}
		\frac{m^2_{\nu_R,ij}(T)}{T^2}&=\frac{1}{8}\left(\tilde{Y}^\dagger_{\nu}\tilde{Y}_{\nu}\right)_{ij}. 
	\end{align}

All the thermal masses are generically non-diagonal due to the possible non-trivial structures of $\tilde{Y}_\ell$ and $\tilde{Y}_{\nu}$. Since these thermal masses appear in the Green functions of finite-temperature field theory, especially in the lepton-doublet thermal propagator element~\cite{Giudice:2003jh},
	\begin{align}\label{thermal prop}
		G_{L}(p)&=\frac{i}{p^2-m^2_{L,ij}(T)+i \epsilon}
		\nonumber \\
		&- 2\pi f_{L}(\vert p^0 \vert)\delta(p^2-m^2_{L,ij}(T)),
	\end{align}
as well as in the approximate dispersion relation $p^2=m^2(T)$, we should firstly go to the \textit{diagonal thermal mass basis} so as to make the finite-temperature calculation well behaved with definite external states. This can be achieved by performing the field redefinition, $L\to V_L^\dagger \hat{L}$, $e_R\to V_R^{e,\dagger} \hat{e}_R$, and $\nu_R \to V_R^{\nu,\dagger} \hat{\nu}_R$, leading then to the following transformed Yukawa matrices:
\begin{align}\label{Ydia}
	\tilde{Y}_\ell\to V_L\tilde{Y}_\ell V_R^{e,\dagger}=\hat{Y}_{\ell}, \quad
	\tilde{Y}_\nu\to V_L\tilde{Y}_\nu V_R^{\nu,\dagger}\equiv Y_{\nu},
\end{align}
with (non-)diagonal $\hat{Y}_{\ell}~({Y}_\nu)$. The above equivalently unitary basis transformation can ensure that all the thermal fermion mass matrices in Eqs.~\eqref{ther-fer-m-2}-\eqref{ther-rfer-m-22} are consistently diagonal in the thermal mass basis. Especially, to visualize the diagonal form of Eq.~\eqref{ther-rfer-m-22}, we can express $\tilde{Y}^\dagger_{\nu}\tilde{Y}_{\nu}$ as $Y^\dagger_{\nu}Y_{\nu}=V_R^{\nu}(\tilde{Y}^\dagger_{\nu}\tilde{Y}_{\nu})V_R^{\nu,\dagger}$ under the unitary basis transformation specified by Eq.~\eqref{Ydia}. Now, it is clear that the Hermitian matrix $\tilde{Y}^\dagger_{\nu}\tilde{Y}_{\nu}$ can always be diagonalized by the unitary matrix $V_R^{\nu}$, and thus Eq.~\eqref{ther-rfer-m-22} is also diagonal once expressed in terms of the product $Y^\dagger_{\nu}Y_{\nu}$ in the transformed basis.

After gauge symmetry breaking, since $\hat{e}_{L,R}$ and $\hat{\nu}_{R}$ are already in the \textit{diagonal vacuum mass basis}, one only needs to transform the left-handed Dirac neutrinos to the vacuum mass basis. To this end, we can formally parametrize the diagonalization as
\begin{align}\label{Ynudia2}
	\hat{Y}_\nu\equiv (U_\nu V_L)\tilde{Y}_\nu V_R^{\nu,\dagger}=U_\nu Y_\nu,
\end{align}
where $U_\nu$ is unitary, and the product $U_\nu V_L$ plays the role of left-unitary rotation.
By simple parameter counting, the diagonalization in Eq.~\eqref{Ynudia2} is always possible via three independent unitary matrices $U_{\nu}$, $V_L$ and $V_R^{\nu}$. It is then clear from Eq.~\eqref{Ynudia2} that $U_\nu$ serves as the desired rotation for the left-handed Dirac neutrinos. Finally, the physical PMNS matrix is simply given by $U_{\mathrm{PMNS}}=U_\nu^\dagger$, and thus
\begin{align}\label{BAU-PMNS portal}
	Y_\nu=\sqrt{2} U_{\mathrm{PMNS}} m_\nu/v_\Phi.
\end{align} 
It should be emphasized that, unlike the common practice of assuming specific Yukawa structures, Eq.~\eqref{BAU-PMNS portal} is a natural consequence of unitary transformations from the original basis in Eq.~\eqref{lag} to the diagonal thermal and vacuum mass bases. In particular, it is the diagonal thermal mass basis that guides the fermion field redefinition and decides which Yukawa set (i.e., $\tilde{Y}_\ell$) should be diagonalized in the finite-temperature regime. If we instead rotate $\tilde{Y}_\nu$ in Eq.~\eqref{lag} to the diagonal basis, the thermal mass matrices $m^2_{L,ij}(T)$ in Eq.~\eqref{ther-fer-m-2} would be non-diagonal and the subsequent $m^2_{L,ij}(T)$-dependent calculation would necessarily involve the non-commutative and complicated matrix algebra, although the thermal basis in Eq.~\eqref{Ydia} and the one introduced to diagonalize $\tilde{Y}_\nu$ are unitarily equivalent. Thus, the importance of diagonal thermal mass basis is that it just assists us, in a much clearer way, to understand the connection between the finite-temperature Yukawa interactions and the low-energy lepton masses and mixing. 

\textit{$\nu_1$-leptogenesis.}--We are now working in the finite-temperature field theory, where the scalar-fermion interactions of Eq.~\eqref{lag} are determined in the diagonal  thermal mass basis. In general, the leptonic \textit{CP} asymmetry ($\epsilon_{CP}$) can be generated by thermal cuts on the self-energy diagrams of $\nu_R$ and/or $L$ from the scalar decay $\bar\Phi\to L \bar \nu_{R}$~\cite{Giudice:2003jh,Hambye:2016sby,Li:2020ner}. However, after a careful analysis of all the possible diagrams and thermal cuts (see, e.g., Ref.~\cite{Li:2020ner} for further technical details), we find the following observations: (i) For the contribution from neutrino self-energy diagram, we find that the resulting $\epsilon_{CP}$ exhibits a very weak dependence on the flavor indices of $L$ propagators and outgoing states. The final leptonic \textit{CP} asymmetry is then obtained trivially by summing over all the $L$-flavor indices from the Yukawa matrices, yielding therefore $\epsilon_{CP}\propto \text{Im}(\text{diag}[\mathcal{H}])=0$, where $\mathcal{H}$ denotes a Hermitian matrix function of $Y_\nu$. (ii) For the self-energy correction in the lepton-doublet line, the right-handed neutrinos (but not charged leptons, since $\hat{Y}_{\ell}$ is already diagonal) propagating in the loop could generate a non-vanishing leptonic \textit{CP} asymmetry. However, if all the three Dirac neutrinos establish the late LRE after the sphaleron process freezes out, which means that the final resulting $\epsilon_{CP}$ should be summed over all the three neutrino flavors, we find that $\epsilon_{CP}$ will be of $\mathcal{O}( Y_\nu^2/\hat{Y}_\ell^2)$, implying  consequently a numerically negligible amount since all the eigenvalues of $Y_\nu$ are much smaller than that of $\hat{Y}_\ell$ in this case. Therefore, (iii) a purely thermal Dirac leptogenesis can become significant only when the leptonic \textit{CP} asymmetry is generated with $\nu_R$ running in the $L$ self-energy loop and \textit{at most} two of the three Dirac neutrinos are allowed to establish the late LRE after the sphaleron process freezes out. 

\begin{figure}[t]
	\centering	
	\includegraphics[width=0.45\textwidth]{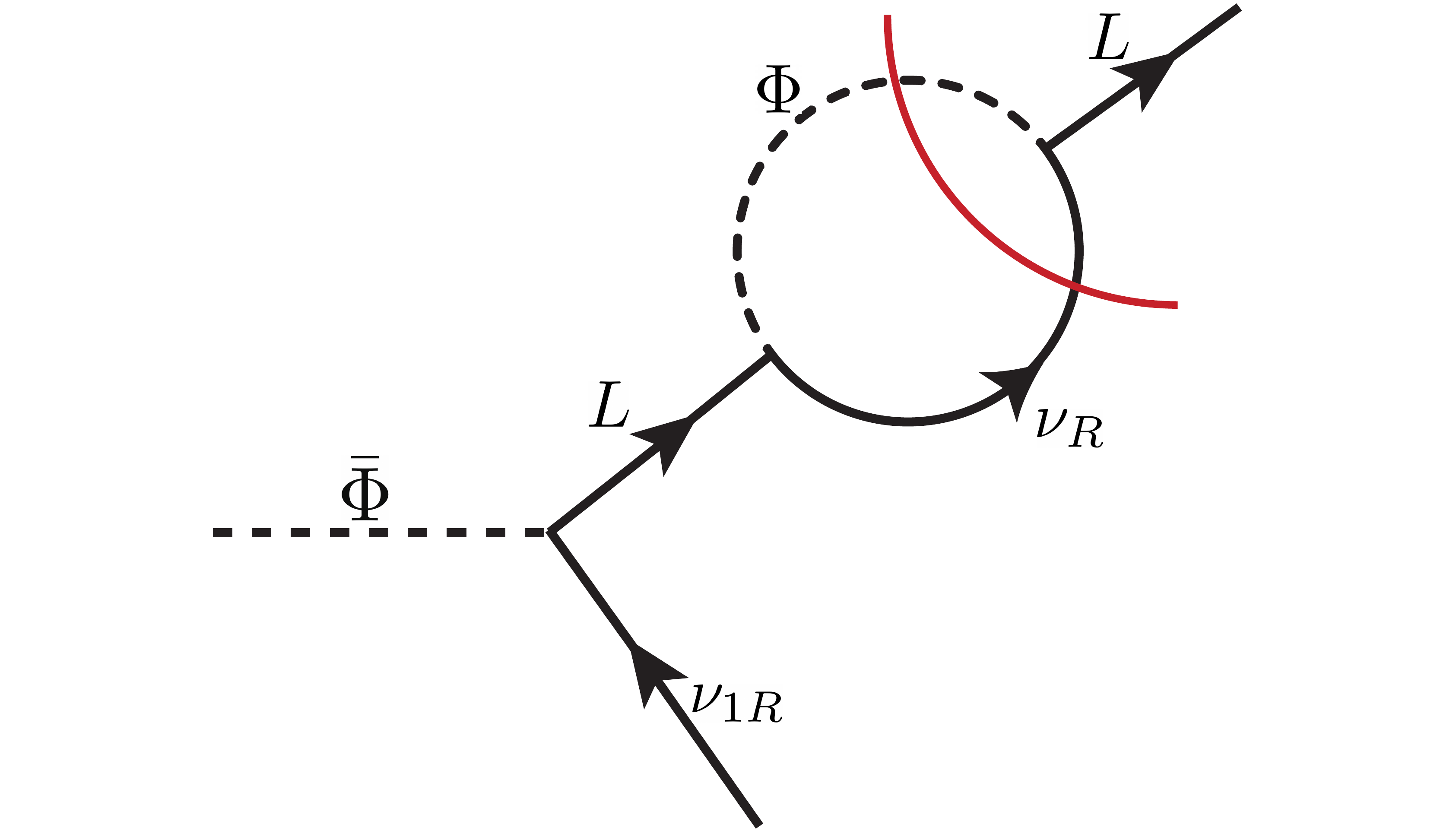}
	\caption{\label{Self-L} Purely thermal leptonic \textit{CP} asymmetry arising from the thermal cut in $\bar\Phi\to L \bar \nu_{1R}$ decay.}
\end{figure}

As a benchmark scenario, it is sufficient to make the lightest Dirac neutrino (denoted by $\nu_1$ and will be explained later) out of equilibrium throughout the sphaleron-active epoch. In this setup, we can immediately obtain a restricted range of the vacuum scale $v_\Phi$ at the sphaleron decoupling temperature $T_{sph}=131.7$~GeV via the LRE conditions for the thermal decay rates:
\begin{align}\label{LREcon}
\langle\Gamma(\Phi\to \bar L \nu_{1R})\rangle\lesssim H(T_{sph}) \lesssim
\langle\Gamma(\Phi\to \bar L \nu_{2(3)R})\rangle, 
\end{align} 
 where the Hubble expansion is given  by $H(T)= T^2/M_{Pl}^*$, with $M_{Pl}^*\equiv M_{Pl}/1.66\sqrt{g_*^\rho}\approx 10^{18}$~GeV. A successful accumulation of leptonic \textit{CP} asymmetry then arises from the decay $\bar{\Phi}\to L \bar{\nu}_{1R}$, as shown in Fig.~\ref{Self-L}. 
 
 To calculate $\epsilon_{CP}$, we adopt the real-time formalism with retarded propagators in thermal field theory~\cite{Kobes:1990ua,Giudice:2003jh,Garny:2010nj,Li:2020ner}, and obtain  \begin{align}\label{CPasym}
	\epsilon_{CP}&\equiv \frac{\Gamma(\Phi\to \bar L \nu_{1R})-\Gamma(\bar\Phi\to L \bar\nu_{1R})}{\Gamma(\Phi\to \bar L \nu_{1R})+\Gamma(\bar\Phi\to L \bar\nu_{1R})}
	\nonumber \\[0.2cm]
	&=\frac{1}{4\pi}\frac{\sum_{i\neq k}\text{Im}[Y_{\nu,i1}^* Y_{\nu,k1}(Y_\nu Y_\nu^\dagger)_{ik}] \mathcal{F}_{ik}}{\sum_j \vert Y_{\nu,j1}\vert^2 (M_\Phi^2-M_{L,j}^2)},
\end{align}
with the self-energy loop function $\mathcal{F}_{ik}$ given by
\begin{align}
\mathcal{F}_{ik} =& \int_{\omega_{min} }^{\omega_{max}}d\omega \frac{M_\Phi^2\left[f_F(-\omega)+f_B(E_i-\omega)\right]}{(M_\Phi^2-M_{L,i}^2)(M_{L,i}^2-M_{L,k}^2)}\nonumber \\[0.2mm]
&\times \Big [ (M_{L,i}^2-M_\Phi^2)M_\Phi-2M_{L,i}^2\omega \Big],
\end{align}
where $\omega_{min}=M_\Phi (M_{L,i}^2-M_\Phi^2)/2M_{L,i}^2$, $\omega_{max}=(M_{L,i}^2-M_\Phi^2)/2M_\Phi$, and $f_{F(B)}(x)=(e^{x/T}\pm1)^{-1}$ is the Fermi-Dirac (Bose-Einstein) distribution. Remarkably, Eq.~\eqref{CPasym} exhibits a fresh dependence on the flavor species, such that the \textit{CP} information is now encoded in the Dirac \textit{CP} phase $\delta_{CP}$. This can be seen from the fact that, in the diagonal thermal mass basis, $Y_\nu$ connects the Dirac neutrino masses ($m_\nu$) with the PMNS matrix via Eq.~\eqref{BAU-PMNS portal}. From Eq.~\eqref{CPasym}, we can also infer that the resulting \textit{CP} asymmetry exhibits a scaling $\epsilon_{CP}\sim Y_\nu^2/\hat{Y}_{\ell}^2$. However, the purely thermal Dirac $\nu_1$-leptogenesis proposed here dramatically differs from the case--corresponding to (ii) observed above--where all the three neutrinos establish the LRE after the sphaleron decoupling. The key point here is that our mechanism predicts relatively larger eigenvalues of $Y_\nu$ to partially enhance the scaling $Y_\nu^2/\hat{Y}_{\ell}^2$. In addition, the smallest entries encoded in $Y_{\nu,i1}$ are canceled out in Eq.~\eqref{CPasym}. This is a realization of the parameter enhancement mechanism (hierarchy of couplings) illustrated in Ref.~\cite{Hambye:2001eu}. 

Before the sphaleron freezes out, all the charged leptons and the two heavier Dirac neutrinos have already established LRE through their Yukawa interactions. Thus, the lepton asymmetry stored in these species would undergo rapid washout processes, resulting therefore in negligible amount of final baryon asymmetry. Nevertheless, a successive accumulation of lepton asymmetry can be stored in the lightest $\nu_R$ part due to its much smaller Yukawa coupling, with the evolution dominated by the simplified freeze-in Boltzmann equation
\begin{align}\label{neutrino asym}
	\frac{dY_{\Delta \nu_{1R}}}{dT}=-\frac{g_\Phi M_\Phi^2}{sH\pi^2}\epsilon_{CP} K_1(M_\Phi/T) \Gamma(\bar\Phi\to  L \bar\nu_{1R}),
\end{align}
where $g_\Phi=2$ results from the two gauge components of $\Phi$, and the SM entropy density is given by $s\approx 48.6 T^3$, while $K_1$ denotes the first modified Bessel function of the second kind. After the sphaleron freezes out, the final baryon asymmetry is fixed by the conversion~\cite{Dick:1999je}
\begin{align}\label{L-to-B}
Y_{\Delta B}=c Y_{\Delta \nu_{1R}},
\end{align}
with $c=8/23$~\cite{Harvey:1990qw}.  

In order to obtain the numerical value of $Y_{\Delta B}$ via Eq.~\eqref{L-to-B}, we simply apply the thermal mass $M_{\Phi}(T)\simeq \sqrt{3g_2^2+g_1^2}/4\,T\simeq 0.3\,T$~\cite{Cline:1995dg} to replace the mass parameter $M_{\Phi}$ in the formulae derived above. Such a treatment can be regarded as an approximation where the mass term from $m_{22}^2 \Phi^\dagger \Phi$ in Eq.~\eqref{pot} is small or even vanishes (say, due to some classical scale symmetry). Note that, even with a small $m_{22}$, the physical masses of the $\Phi$ components at the electroweak scale can still be generated by the SM-like Higgs vacuum via the $\lambda_3$ term in Eq.~\eqref{pot}, as can be seen from Eq.~\eqref{eq:massspec} and the  discussion below it.
 
In our numerical setup, the vacuum scale satisfying Eq.~\eqref{LREcon} with a normal-ordering neutrino mass spectrum $m_1<m_2<m_3$~\cite{Zyla:2020zbs} is predicted to be
\begin{align}\label{vPhi-range}
\left(\frac{m_1}{\rm meV}\right)\lesssim \left(\frac{v_\Phi}{1.31\; \rm keV}\right)\lesssim \sqrt{\left(\frac{m_1}{\rm meV}\right)^2+73.90}.
\end{align} 
Note that, as $dY_{\Delta \nu_{1R}}/dT\propto 1/T^2$, the temperature integration is insensitive to the initial value $T_{i}$ when the sphaleron comes into thermalization around $T_{i}=10^{12}$~GeV. On the other hand, we apply the critical temperature of gauge symmetry breaking, $T_{c}\approx 160$~GeV, as the lower integration limit. This is reasonable because the generation rate of lepton asymmetry quickly becomes Boltzmann suppressed after the scalar obtains an electroweak vacuum mass from phase transition. In this context, integrating Eq.~\eqref{neutrino asym} over the temperature will result in a simple expression of the baryon asymmetry:
\begin{align}\label{YBnum}
	Y_{\Delta B}&\approx Y_{\Delta B}^{\mathrm{exp}} \left(\frac{m_1}{\mathrm{meV}}\right)^2\left(\frac{v_\Phi}{5.18\;\mathrm{keV}}\right)^{-4} \left(\frac{\sin\delta_{CP}}{0.64}\right),
\end{align}
where $Y_{\Delta B}^{\rm exp}\approx 8.75\times 10^{-11}$ corresponds to the currently observed baryon asymmetry~\cite{Aghanim:2018eyx}, and the best-fit mixing angles of $U_{\mathrm{PMNS}}$ as well as the normal-ordering neutrino mass-squared differences have been used~\cite{Zyla:2020zbs}.  

Eq.~\eqref{YBnum} displays a clear connection between the finite-temperature generated baryon asymmetry and the low-energy leptonic \textit{CP} violation. A confirming detection of the Dirac \textit{CP} phase with $\delta_{CP}\neq 0,\pi$ in the upcoming experiments can therefore unambiguously support the purely thermal Dirac $\nu_1$-leptogenesis proposed here. In the particular case where the bound in Eq.~\eqref{vPhi-range} is saturated (i.e., either the lightest or the second lightest Dirac neutrino happens to establish LRE around $T_{sph}$), the amount of baryon asymmetry can then be estimated by the lightest neutrino mass and the Dirac \textit{CP} phase. Noticeably in this case, a small \textit{CP}-violating phase of $\mathcal{O}(10^{-3})$ can still account for the observed baryon asymmetry as long as the lightest neutrino mass is of $\mathcal{O}(1)$~meV, which is well compatible with the cosmological bound $\sum_i m_i<0.12$~eV~\cite{Aghanim:2018eyx} (see also Ref.~\cite{Vagnozzi:2017ovm}). 

\begin{figure}[t]
	\centering	
	\includegraphics[width=0.45\textwidth]{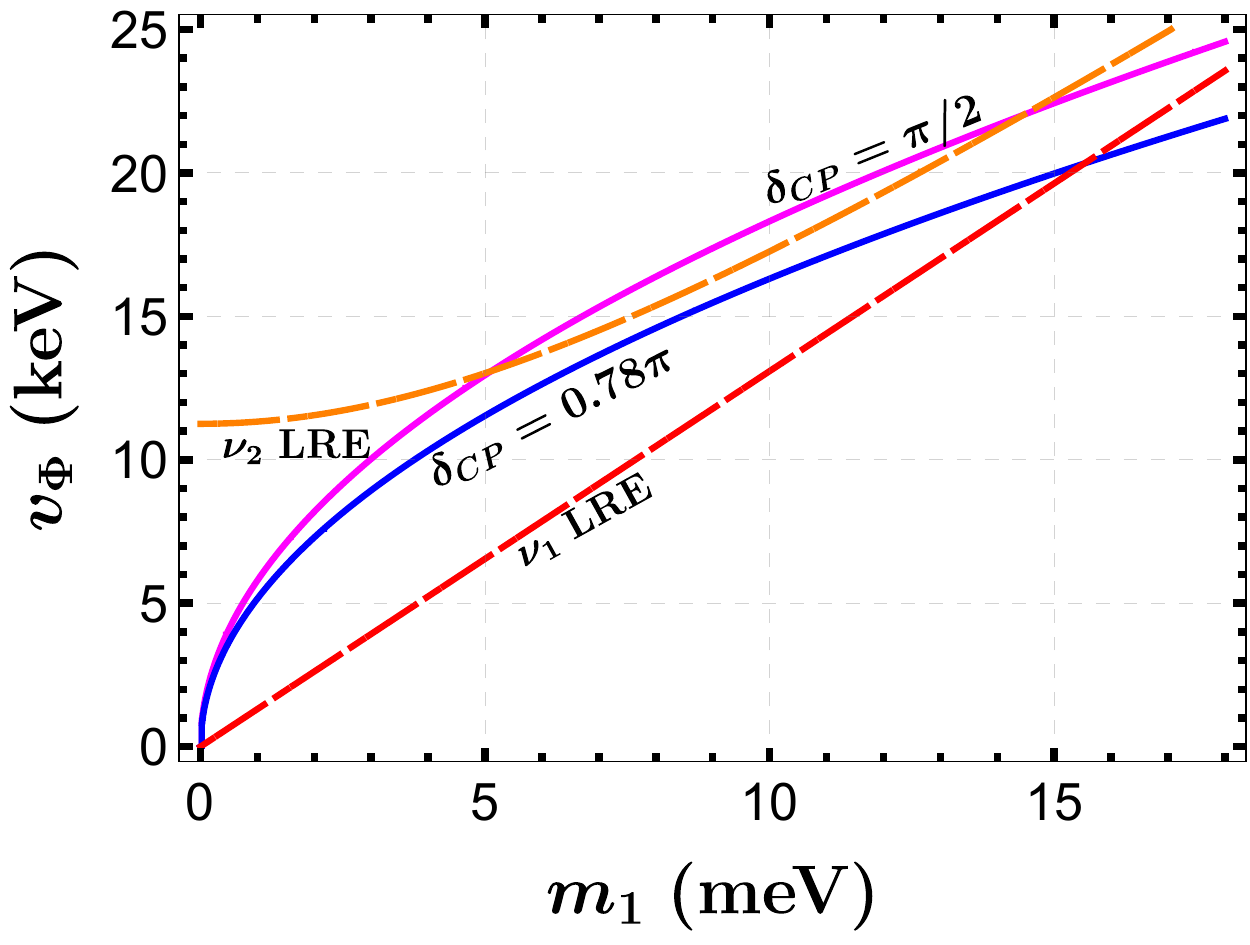}
	\caption{\label{Bfit} Correlation between the lightest neutrino mass $m_1$ and the vacuum scale $v_{\Phi}$ in generating the observed baryon asymmetry $Y_{\Delta B}^{\rm exp}=8.75\times 10^{-11}$~\cite{Aghanim:2018eyx}. See the text for details.}
\end{figure}

In Fig.~\ref{Bfit}, we show the correlation between the lightest neutrino mass $m_1$ and the vacuum scale $v_{\Phi}$ in generating the observed baryon asymmetry $Y_{\Delta B}^{\rm exp}=8.75\times 10^{-11}$~\cite{Aghanim:2018eyx}. Here the magenta curve corresponds to the case with maximal \textit{CP} violation $\delta_{CP}=\pi/2$, and the region below the red (orange) dashed curve signifies the LRE of the lightest (the second lightest) Dirac neutrino at $T_{sph}$. As a comparison, we have also considered the case with the $3\sigma$ lower bound $\delta_{CP}\geqslant 0.78\pi$ obtained by combining the global fitting results of three-neutrino oscillation data~\cite{Zyla:2020zbs}, which is shown by the blue solid curve. In this case, the upper bound on the lightest Dirac neutrino mass is constrained to be around $15$~meV. It should be mentioned that, with an $\mathcal{O}(10)$~keV vacuum $v_\Phi$ and an meV-scale $\nu_1$, we can predict an $\mathcal{O}(10^{-7})$ eigenvalue for the smallest Dirac neutrino Yukawa, while an order of the electron Yukawa, $Y_e\simeq\mathcal{O}(10^{-6}) $, for the two larger ones. As a consequence, the approximation based on Eq.~\eqref{ther-fer-m-2} is numerically justified.  

Note that the discussion made thus far is based on a normal-ordering neutrino mass spectrum. As the neutrino mass ordering hinted so far has a close relation with the range of the Dirac \textit{CP} phase (see e.g., Refs.~\cite{Abe:2019vii,Acero:2019ksn}), the purely thermal Dirac $\nu_1$-leptogenesis can also help to infer the neutrino mass ordering. In fact, if the currently averaged $3\sigma$ lower bound on the Dirac \textit{CP} phase is considered~\cite{Zyla:2020zbs}, the inverted-ordering pattern, $m_3<m_1<m_2$, cannot create a Universe with matter dominating over antimatter, because now $\delta_{CP}>\pi$. Therefore, the baryogenesis from the lightest Dirac neutrino favors a normal-ordering mass spectrum in light of the current data, and this is the reason why the benchmark model is dubbed \textit{$\nu_1$-leptogenesis}.

Let us finally comment on some possible probes of the $\nu_1$-leptogenesis. Under the minimal setup with Eqs.~\eqref{lag} and \eqref{pot}, the scenario mimics a neutrinophilic two-Higgs-doublet model~\cite{Davidson:2009ha}. However, the $\nu_1$-leptogenesis predicts an $\mathcal{O}(10)$~keV vacuum $v_\Phi$, which is different from the earlier studies where order-one Dirac neutrino Yukawa couplings and eV-scale vacuum $v_\Phi$ were focused on. Following the studies made in Refs.~\cite{Machado:2015sha,Bertuzzo:2015ada} as well as the discussion in Ref.~\cite{Li:2020ner}, it can be found that constraints from the low-energy flavor physics, such as the lepton-flavor violating processes, and the collider physics, such as the quark-associated and dilepton decay products, are diluted in the $\nu_1$-leptogenesis, due to the feeble Dirac neutrino Yukawa couplings as well as the neutrinophilic non-SM Higgs bosons. Nevertheless, the relativistic right-handed Dirac neutrinos in the early Universe can act as extra radiation beyond the standard cosmological model, and hence potentially cause observable effects through cosmological detection~\cite{Li-2022}.

\textit{Conclusion.}--In summary, we have applied in this Letter the thermal field theory to the Dirac leptogenesis. Guided naturally by the diagonal thermal mass basis in the finite-temperature regime, we obtain a clear connection between the low-energy observables and the quantities participating in the finite-temperature baryon asymmetry generation. Under the finite-temperature circumstance, we have presented a purely thermal Dirac $\nu_1$-leptogenesis that formulates the baryon asymmetry in terms of the masses and mixing from hierarchical Dirac neutrinos, together with a restricted vacuum scale accountable for the smallness of Dirac neutrino masses via a seesaw-like relation. The scenario allows us to establish the unique phase source without particular assumptions and/or flavor symmetries on the unknown Yukawa structures. The minimal scenario presented here can be applied as a prototype to a broad class of well-motivated model buildings, and it can also be tested by future neutrino oscillation experiments and possible cosmological detection.
 
This work is supported by the National Natural Science Foundation of China under Grant Nos.~12075097, 12047527, 11675061 and 11775092, as well as by the Fundamental Research Funds for the Central Universities under Grant Nos.~CCNU20TS007 and~2020YBZZ074.

\bibliographystyle{apsrev4-1}
\bibliography{reference}

\begin{thebibliography}{40}%
\makeatletter
\providecommand \@ifxundefined [1]{%
 \@ifx{#1\undefined}
}%
\providecommand \@ifnum [1]{%
 \ifnum #1\expandafter \@firstoftwo
 \else \expandafter \@secondoftwo
 \fi
}%
\providecommand \@ifx [1]{%
 \ifx #1\expandafter \@firstoftwo
 \else \expandafter \@secondoftwo
 \fi
}%
\providecommand \natexlab [1]{#1}%
\providecommand \enquote  [1]{``#1''}%
\providecommand \bibnamefont  [1]{#1}%
\providecommand \bibfnamefont [1]{#1}%
\providecommand \citenamefont [1]{#1}%
\providecommand \href@noop [0]{\@secondoftwo}%
\providecommand \href [0]{\begingroup \@sanitize@url \@href}%
\providecommand \@href[1]{\@@startlink{#1}\@@href}%
\providecommand \@@href[1]{\endgroup#1\@@endlink}%
\providecommand \@sanitize@url [0]{\catcode `\\12\catcode `\$12\catcode
  `\&12\catcode `\#12\catcode `\^12\catcode `\_12\catcode `\%12\relax}%
\providecommand \@@startlink[1]{}%
\providecommand \@@endlink[0]{}%
\providecommand \url  [0]{\begingroup\@sanitize@url \@url }%
\providecommand \@url [1]{\endgroup\@href {#1}{\urlprefix }}%
\providecommand \urlprefix  [0]{URL }%
\providecommand \Eprint [0]{\href }%
\providecommand \doibase [0]{http://dx.doi.org/}%
\providecommand \selectlanguage [0]{\@gobble}%
\providecommand \bibinfo  [0]{\@secondoftwo}%
\providecommand \bibfield  [0]{\@secondoftwo}%
\providecommand \translation [1]{[#1]}%
\providecommand \BibitemOpen [0]{}%
\providecommand \bibitemStop [0]{}%
\providecommand \bibitemNoStop [0]{.\EOS\space}%
\providecommand \EOS [0]{\spacefactor3000\relax}%
\providecommand \BibitemShut  [1]{\csname bibitem#1\endcsname}%
\let\auto@bib@innerbib\@empty
\bibitem [{\citenamefont {Fukugita}\ and\ \citenamefont
  {Yanagida}(1986)}]{Fukugita:1986hr}%
  \BibitemOpen
  \bibfield  {author} {\bibinfo {author} {\bibfnamefont {M.}~\bibnamefont
  {Fukugita}}\ and\ \bibinfo {author} {\bibfnamefont {T.}~\bibnamefont
  {Yanagida}},\ }\href {\doibase 10.1016/0370-2693(86)91126-3} {\bibfield
  {journal} {\bibinfo  {journal} {Phys. Lett. B}\ }\textbf {\bibinfo {volume}
  {174}},\ \bibinfo {pages} {45} (\bibinfo {year} {1986})}\BibitemShut
  {NoStop}%
\bibitem [{\citenamefont {Buchmuller}\ and\ \citenamefont
  {Plumacher}(1996)}]{Buchmuller:1996pa}%
  \BibitemOpen
  \bibfield  {author} {\bibinfo {author} {\bibfnamefont {W.}~\bibnamefont
  {Buchmuller}}\ and\ \bibinfo {author} {\bibfnamefont {M.}~\bibnamefont
  {Plumacher}},\ }\href {\doibase 10.1016/S0370-2693(96)01232-4} {\bibfield
  {journal} {\bibinfo  {journal} {Phys. Lett. B}\ }\textbf {\bibinfo {volume}
  {389}},\ \bibinfo {pages} {73} (\bibinfo {year} {1996})},\ \Eprint
  {http://arxiv.org/abs/hep-ph/9608308} {arXiv:hep-ph/9608308} \BibitemShut
  {NoStop}%
\bibitem [{\citenamefont {Branco}\ \emph {et~al.}(2001)\citenamefont {Branco},
  \citenamefont {Morozumi}, \citenamefont {Nobre},\ and\ \citenamefont
  {Rebelo}}]{Branco:2001pq}%
  \BibitemOpen
  \bibfield  {author} {\bibinfo {author} {\bibfnamefont {G.~C.}\ \bibnamefont
  {Branco}}, \bibinfo {author} {\bibfnamefont {T.}~\bibnamefont {Morozumi}},
  \bibinfo {author} {\bibfnamefont {B.}~\bibnamefont {Nobre}}, \ and\ \bibinfo
  {author} {\bibfnamefont {M.}~\bibnamefont {Rebelo}},\ }\href {\doibase
  10.1016/S0550-3213(01)00425-4} {\bibfield  {journal} {\bibinfo  {journal}
  {Nucl. Phys. B}\ }\textbf {\bibinfo {volume} {617}},\ \bibinfo {pages} {475}
  (\bibinfo {year} {2001})},\ \Eprint {http://arxiv.org/abs/hep-ph/0107164}
  {arXiv:hep-ph/0107164} \BibitemShut {NoStop}%
\bibitem [{\citenamefont {Branco}\ \emph {et~al.}(2002)\citenamefont {Branco},
  \citenamefont {Gonzalez~Felipe}, \citenamefont {Joaquim},\ and\ \citenamefont
  {Rebelo}}]{Branco:2002kt}%
  \BibitemOpen
  \bibfield  {author} {\bibinfo {author} {\bibfnamefont {G.}~\bibnamefont
  {Branco}}, \bibinfo {author} {\bibfnamefont {R.}~\bibnamefont
  {Gonzalez~Felipe}}, \bibinfo {author} {\bibfnamefont {F.}~\bibnamefont
  {Joaquim}}, \ and\ \bibinfo {author} {\bibfnamefont {M.}~\bibnamefont
  {Rebelo}},\ }\href {\doibase 10.1016/S0550-3213(02)00478-9} {\bibfield
  {journal} {\bibinfo  {journal} {Nucl. Phys. B}\ }\textbf {\bibinfo {volume}
  {640}},\ \bibinfo {pages} {202} (\bibinfo {year} {2002})},\ \Eprint
  {http://arxiv.org/abs/hep-ph/0202030} {arXiv:hep-ph/0202030} \BibitemShut
  {NoStop}%
\bibitem [{\citenamefont {Akhmedov}\ \emph {et~al.}(1998)\citenamefont
  {Akhmedov}, \citenamefont {Rubakov},\ and\ \citenamefont
  {Smirnov}}]{Akhmedov:1998qx}%
  \BibitemOpen
  \bibfield  {author} {\bibinfo {author} {\bibfnamefont {E.~K.}\ \bibnamefont
  {Akhmedov}}, \bibinfo {author} {\bibfnamefont {V.}~\bibnamefont {Rubakov}}, \
  and\ \bibinfo {author} {\bibfnamefont {A.}~\bibnamefont {Smirnov}},\ }\href
  {\doibase 10.1103/PhysRevLett.81.1359} {\bibfield  {journal} {\bibinfo
  {journal} {Phys. Rev. Lett.}\ }\textbf {\bibinfo {volume} {81}},\ \bibinfo
  {pages} {1359} (\bibinfo {year} {1998})},\ \Eprint
  {http://arxiv.org/abs/hep-ph/9803255} {arXiv:hep-ph/9803255} \BibitemShut
  {NoStop}%
\bibitem [{\citenamefont {Pilaftsis}\ and\ \citenamefont
  {Underwood}(2004)}]{Pilaftsis:2003gt}%
  \BibitemOpen
  \bibfield  {author} {\bibinfo {author} {\bibfnamefont {A.}~\bibnamefont
  {Pilaftsis}}\ and\ \bibinfo {author} {\bibfnamefont {T.~E.}\ \bibnamefont
  {Underwood}},\ }\href {\doibase 10.1016/j.nuclphysb.2004.05.029} {\bibfield
  {journal} {\bibinfo  {journal} {Nucl. Phys. B}\ }\textbf {\bibinfo {volume}
  {692}},\ \bibinfo {pages} {303} (\bibinfo {year} {2004})},\ \Eprint
  {http://arxiv.org/abs/hep-ph/0309342} {arXiv:hep-ph/0309342} \BibitemShut
  {NoStop}%
\bibitem [{\citenamefont {Hambye}\ and\ \citenamefont
  {Teresi}(2016)}]{Hambye:2016sby}%
  \BibitemOpen
  \bibfield  {author} {\bibinfo {author} {\bibfnamefont {T.}~\bibnamefont
  {Hambye}}\ and\ \bibinfo {author} {\bibfnamefont {D.}~\bibnamefont
  {Teresi}},\ }\href {\doibase 10.1103/PhysRevLett.117.091801} {\bibfield
  {journal} {\bibinfo  {journal} {Phys. Rev. Lett.}\ }\textbf {\bibinfo
  {volume} {117}},\ \bibinfo {pages} {091801} (\bibinfo {year} {2016})},\
  \Eprint {http://arxiv.org/abs/1606.00017} {arXiv:1606.00017 [hep-ph]}
  \BibitemShut {NoStop}%
\bibitem [{\citenamefont {Casas}\ and\ \citenamefont
  {Ibarra}(2001)}]{Casas:2001sr}%
  \BibitemOpen
  \bibfield  {author} {\bibinfo {author} {\bibfnamefont {J.}~\bibnamefont
  {Casas}}\ and\ \bibinfo {author} {\bibfnamefont {A.}~\bibnamefont {Ibarra}},\
  }\href {\doibase 10.1016/S0550-3213(01)00475-8} {\bibfield  {journal}
  {\bibinfo  {journal} {Nucl. Phys. B}\ }\textbf {\bibinfo {volume} {618}},\
  \bibinfo {pages} {171} (\bibinfo {year} {2001})},\ \Eprint
  {http://arxiv.org/abs/hep-ph/0103065} {arXiv:hep-ph/0103065} \BibitemShut
  {NoStop}%
\bibitem [{\citenamefont {Buchmuller}\ and\ \citenamefont
  {Plumacher}(2000)}]{Buchmuller:2000as}%
  \BibitemOpen
  \bibfield  {author} {\bibinfo {author} {\bibfnamefont {W.}~\bibnamefont
  {Buchmuller}}\ and\ \bibinfo {author} {\bibfnamefont {M.}~\bibnamefont
  {Plumacher}},\ }\href {\doibase 10.1016/S0217-751X(00)00293-5} {\bibfield
  {journal} {\bibinfo  {journal} {Int. J. Mod. Phys. A}\ }\textbf {\bibinfo
  {volume} {15}},\ \bibinfo {pages} {5047} (\bibinfo {year} {2000})},\ \Eprint
  {http://arxiv.org/abs/hep-ph/0007176} {arXiv:hep-ph/0007176} \BibitemShut
  {NoStop}%
\bibitem [{\citenamefont {Joshipura}\ \emph {et~al.}(2001)\citenamefont
  {Joshipura}, \citenamefont {Paschos},\ and\ \citenamefont
  {Rodejohann}}]{Joshipura:2001ui}%
  \BibitemOpen
  \bibfield  {author} {\bibinfo {author} {\bibfnamefont {A.~S.}\ \bibnamefont
  {Joshipura}}, \bibinfo {author} {\bibfnamefont {E.~A.}\ \bibnamefont
  {Paschos}}, \ and\ \bibinfo {author} {\bibfnamefont {W.}~\bibnamefont
  {Rodejohann}},\ }\href {\doibase 10.1088/1126-6708/2001/08/029} {\bibfield
  {journal} {\bibinfo  {journal} {JHEP}\ }\textbf {\bibinfo {volume} {08}},\
  \bibinfo {pages} {029} (\bibinfo {year} {2001})},\ \Eprint
  {http://arxiv.org/abs/hep-ph/0105175} {arXiv:hep-ph/0105175} \BibitemShut
  {NoStop}%
\bibitem [{\citenamefont {Davidson}\ and\ \citenamefont
  {Ibarra}(2001)}]{Davidson:2001zk}%
  \BibitemOpen
  \bibfield  {author} {\bibinfo {author} {\bibfnamefont {S.}~\bibnamefont
  {Davidson}}\ and\ \bibinfo {author} {\bibfnamefont {A.}~\bibnamefont
  {Ibarra}},\ }\href {\doibase 10.1088/1126-6708/2001/09/013} {\bibfield
  {journal} {\bibinfo  {journal} {JHEP}\ }\textbf {\bibinfo {volume} {09}},\
  \bibinfo {pages} {013} (\bibinfo {year} {2001})},\ \Eprint
  {http://arxiv.org/abs/hep-ph/0104076} {arXiv:hep-ph/0104076} \BibitemShut
  {NoStop}%
\bibitem [{\citenamefont {Moffat}\ \emph {et~al.}(2019)\citenamefont {Moffat},
  \citenamefont {Pascoli}, \citenamefont {Petcov},\ and\ \citenamefont
  {Turner}}]{Moffat:2018smo}%
  \BibitemOpen
  \bibfield  {author} {\bibinfo {author} {\bibfnamefont {K.}~\bibnamefont
  {Moffat}}, \bibinfo {author} {\bibfnamefont {S.}~\bibnamefont {Pascoli}},
  \bibinfo {author} {\bibfnamefont {S.~T.}\ \bibnamefont {Petcov}}, \ and\
  \bibinfo {author} {\bibfnamefont {J.}~\bibnamefont {Turner}},\ }\href
  {\doibase 10.1007/JHEP03(2019)034} {\bibfield  {journal} {\bibinfo  {journal}
  {JHEP}\ }\textbf {\bibinfo {volume} {03}},\ \bibinfo {pages} {034} (\bibinfo
  {year} {2019})},\ \Eprint {http://arxiv.org/abs/1809.08251} {arXiv:1809.08251
  [hep-ph]} \BibitemShut {NoStop}%
\bibitem [{\citenamefont {Xing}\ and\ \citenamefont
  {Zhang}(2020{\natexlab{a}})}]{Xing:2020ghj}%
  \BibitemOpen
  \bibfield  {author} {\bibinfo {author} {\bibfnamefont {Z.-z.}\ \bibnamefont
  {Xing}}\ and\ \bibinfo {author} {\bibfnamefont {D.}~\bibnamefont {Zhang}},\
  }\href {\doibase 10.1016/j.physletb.2020.135397} {\bibfield  {journal}
  {\bibinfo  {journal} {Phys. Lett. B}\ }\textbf {\bibinfo {volume} {804}},\
  \bibinfo {pages} {135397} (\bibinfo {year} {2020}{\natexlab{a}})},\ \Eprint
  {http://arxiv.org/abs/2003.06312} {arXiv:2003.06312 [hep-ph]} \BibitemShut
  {NoStop}%
\bibitem [{\citenamefont {Xing}\ and\ \citenamefont
  {Zhang}(2020{\natexlab{b}})}]{Xing:2020erm}%
  \BibitemOpen
  \bibfield  {author} {\bibinfo {author} {\bibfnamefont {Z.-z.}\ \bibnamefont
  {Xing}}\ and\ \bibinfo {author} {\bibfnamefont {D.}~\bibnamefont {Zhang}},\
  }\href {\doibase 10.1007/JHEP04(2020)179} {\bibfield  {journal} {\bibinfo
  {journal} {JHEP}\ }\textbf {\bibinfo {volume} {04}},\ \bibinfo {pages} {179}
  (\bibinfo {year} {2020}{\natexlab{b}})},\ \Eprint
  {http://arxiv.org/abs/2003.00480} {arXiv:2003.00480 [hep-ph]} \BibitemShut
  {NoStop}%
\bibitem [{\citenamefont {Rahat}(2021)}]{Rahat:2020mio}%
  \BibitemOpen
  \bibfield  {author} {\bibinfo {author} {\bibfnamefont {M.~H.}\ \bibnamefont
  {Rahat}},\ }\href {\doibase 10.1103/PhysRevD.103.035011} {\bibfield
  {journal} {\bibinfo  {journal} {Phys. Rev. D}\ }\textbf {\bibinfo {volume}
  {103}},\ \bibinfo {pages} {035011} (\bibinfo {year} {2021})},\ \Eprint
  {http://arxiv.org/abs/2008.04204} {arXiv:2008.04204 [hep-ph]} \BibitemShut
  {NoStop}%
\bibitem [{\citenamefont {Dick}\ \emph {et~al.}(2000)\citenamefont {Dick},
  \citenamefont {Lindner}, \citenamefont {Ratz},\ and\ \citenamefont
  {Wright}}]{Dick:1999je}%
  \BibitemOpen
  \bibfield  {author} {\bibinfo {author} {\bibfnamefont {K.}~\bibnamefont
  {Dick}}, \bibinfo {author} {\bibfnamefont {M.}~\bibnamefont {Lindner}},
  \bibinfo {author} {\bibfnamefont {M.}~\bibnamefont {Ratz}}, \ and\ \bibinfo
  {author} {\bibfnamefont {D.}~\bibnamefont {Wright}},\ }\href {\doibase
  10.1103/PhysRevLett.84.4039} {\bibfield  {journal} {\bibinfo  {journal}
  {Phys. Rev. Lett.}\ }\textbf {\bibinfo {volume} {84}},\ \bibinfo {pages}
  {4039} (\bibinfo {year} {2000})},\ \Eprint
  {http://arxiv.org/abs/hep-ph/9907562} {arXiv:hep-ph/9907562 [hep-ph]}
  \BibitemShut {NoStop}%
\bibitem [{\citenamefont {Gu}\ and\ \citenamefont {He}(2006)}]{Gu:2006dc}%
  \BibitemOpen
  \bibfield  {author} {\bibinfo {author} {\bibfnamefont {P.-H.}\ \bibnamefont
  {Gu}}\ and\ \bibinfo {author} {\bibfnamefont {H.-J.}\ \bibnamefont {He}},\
  }\href {\doibase 10.1088/1475-7516/2006/12/010} {\bibfield  {journal}
  {\bibinfo  {journal} {JCAP}\ }\textbf {\bibinfo {volume} {12}},\ \bibinfo
  {pages} {010} (\bibinfo {year} {2006})},\ \Eprint
  {http://arxiv.org/abs/hep-ph/0610275} {arXiv:hep-ph/0610275} \BibitemShut
  {NoStop}%
\bibitem [{\citenamefont {Gu}\ \emph {et~al.}(2007)\citenamefont {Gu},
  \citenamefont {He},\ and\ \citenamefont {Sarkar}}]{Gu:2007mi}%
  \BibitemOpen
  \bibfield  {author} {\bibinfo {author} {\bibfnamefont {P.-H.}\ \bibnamefont
  {Gu}}, \bibinfo {author} {\bibfnamefont {H.-J.}\ \bibnamefont {He}}, \ and\
  \bibinfo {author} {\bibfnamefont {U.}~\bibnamefont {Sarkar}},\ }\href
  {\doibase 10.1088/1475-7516/2007/11/016} {\bibfield  {journal} {\bibinfo
  {journal} {JCAP}\ }\textbf {\bibinfo {volume} {11}},\ \bibinfo {pages} {016}
  (\bibinfo {year} {2007})},\ \Eprint {http://arxiv.org/abs/0705.3736}
  {arXiv:0705.3736 [hep-ph]} \BibitemShut {NoStop}%
\bibitem [{\citenamefont {Wang}\ and\ \citenamefont
  {Han}(2017)}]{Wang:2016lve}%
  \BibitemOpen
  \bibfield  {author} {\bibinfo {author} {\bibfnamefont {W.}~\bibnamefont
  {Wang}}\ and\ \bibinfo {author} {\bibfnamefont {Z.-L.}\ \bibnamefont {Han}},\
  }\href {\doibase 10.1007/JHEP04(2017)166} {\bibfield  {journal} {\bibinfo
  {journal} {JHEP}\ }\textbf {\bibinfo {volume} {04}},\ \bibinfo {pages} {166}
  (\bibinfo {year} {2017})},\ \Eprint {http://arxiv.org/abs/1611.03240}
  {arXiv:1611.03240 [hep-ph]} \BibitemShut {NoStop}%
\bibitem [{\citenamefont {Li}\ \emph {et~al.}(2020)\citenamefont {Li},
  \citenamefont {Li}, \citenamefont {Yan},\ and\ \citenamefont
  {Yang}}]{Li:2020ner}%
  \BibitemOpen
  \bibfield  {author} {\bibinfo {author} {\bibfnamefont {S.-P.}\ \bibnamefont
  {Li}}, \bibinfo {author} {\bibfnamefont {X.-Q.}\ \bibnamefont {Li}}, \bibinfo
  {author} {\bibfnamefont {X.-S.}\ \bibnamefont {Yan}}, \ and\ \bibinfo
  {author} {\bibfnamefont {Y.-D.}\ \bibnamefont {Yang}},\ }\href {\doibase
  10.1140/epjc/s10052-020-08696-z} {\bibfield  {journal} {\bibinfo  {journal}
  {Eur. Phys. J. C}\ }\textbf {\bibinfo {volume} {80}},\ \bibinfo {pages}
  {1122} (\bibinfo {year} {2020})},\ \Eprint {http://arxiv.org/abs/2005.02927}
  {arXiv:2005.02927 [hep-ph]} \BibitemShut {NoStop}%
\bibitem [{\citenamefont {Branco}\ \emph {et~al.}(2012)\citenamefont {Branco},
  \citenamefont {Felipe},\ and\ \citenamefont {Joaquim}}]{Branco:2011zb}%
  \BibitemOpen
  \bibfield  {author} {\bibinfo {author} {\bibfnamefont {G.}~\bibnamefont
  {Branco}}, \bibinfo {author} {\bibfnamefont {R.}~\bibnamefont {Felipe}}, \
  and\ \bibinfo {author} {\bibfnamefont {F.}~\bibnamefont {Joaquim}},\ }\href
  {\doibase 10.1103/RevModPhys.84.515} {\bibfield  {journal} {\bibinfo
  {journal} {Rev. Mod. Phys.}\ }\textbf {\bibinfo {volume} {84}},\ \bibinfo
  {pages} {515} (\bibinfo {year} {2012})},\ \Eprint
  {http://arxiv.org/abs/1111.5332} {arXiv:1111.5332 [hep-ph]} \BibitemShut
  {NoStop}%
\bibitem [{\citenamefont {Rebelo}(2003)}]{Rebelo:2002wj}%
  \BibitemOpen
  \bibfield  {author} {\bibinfo {author} {\bibfnamefont {M.}~\bibnamefont
  {Rebelo}},\ }\href {\doibase 10.1103/PhysRevD.67.013008} {\bibfield
  {journal} {\bibinfo  {journal} {Phys. Rev. D}\ }\textbf {\bibinfo {volume}
  {67}},\ \bibinfo {pages} {013008} (\bibinfo {year} {2003})},\ \Eprint
  {http://arxiv.org/abs/hep-ph/0207236} {arXiv:hep-ph/0207236} \BibitemShut
  {NoStop}%
\bibitem [{\citenamefont {Abe}\ \emph {et~al.}(2020)\citenamefont {Abe} \emph
  {et~al.}}]{Abe:2019vii}%
  \BibitemOpen
  \bibfield  {author} {\bibinfo {author} {\bibfnamefont {K.}~\bibnamefont
  {Abe}} \emph {et~al.} (\bibinfo {collaboration} {T2K}),\ }\href {\doibase
  10.1038/s41586-020-2177-0} {\bibfield  {journal} {\bibinfo  {journal}
  {Nature}\ }\textbf {\bibinfo {volume} {580}},\ \bibinfo {pages} {339}
  (\bibinfo {year} {2020})},\ \bibinfo {note} {[Erratum: Nature 583, E16
  (2020)]},\ \Eprint {http://arxiv.org/abs/1910.03887} {arXiv:1910.03887
  [hep-ex]} \BibitemShut {NoStop}%
\bibitem [{\citenamefont {Das}(1997)}]{Das1997}%
  \BibitemOpen
  \bibfield  {author} {\bibinfo {author} {\bibfnamefont {A.}~\bibnamefont
  {Das}},\ }\href {\doibase 10.1142/3277} {\emph {\bibinfo {title} {Finite
  Temperature Field Theory}}}\ (\bibinfo  {publisher} {WORLD SCIENTIFIC},\
  \bibinfo {year} {1997})\ \Eprint
  {http://arxiv.org/abs/https://www.worldscientific.com/doi/pdf/10.1142/3277}
  {https://www.worldscientific.com/doi/pdf/10.1142/3277} \BibitemShut {NoStop}%
\bibitem [{\citenamefont {Giudice}\ \emph {et~al.}(2004)\citenamefont
  {Giudice}, \citenamefont {Notari}, \citenamefont {Raidal}, \citenamefont
  {Riotto},\ and\ \citenamefont {Strumia}}]{Giudice:2003jh}%
  \BibitemOpen
  \bibfield  {author} {\bibinfo {author} {\bibfnamefont {G.~F.}\ \bibnamefont
  {Giudice}}, \bibinfo {author} {\bibfnamefont {A.}~\bibnamefont {Notari}},
  \bibinfo {author} {\bibfnamefont {M.}~\bibnamefont {Raidal}}, \bibinfo
  {author} {\bibfnamefont {A.}~\bibnamefont {Riotto}}, \ and\ \bibinfo {author}
  {\bibfnamefont {A.}~\bibnamefont {Strumia}},\ }\href {\doibase
  10.1016/j.nuclphysb.2004.02.019} {\bibfield  {journal} {\bibinfo  {journal}
  {Nucl. Phys.}\ }\textbf {\bibinfo {volume} {B685}},\ \bibinfo {pages} {89}
  (\bibinfo {year} {2004})},\ \Eprint {http://arxiv.org/abs/hep-ph/0310123}
  {arXiv:hep-ph/0310123 [hep-ph]} \BibitemShut {NoStop}%
\bibitem [{\citenamefont {Davidson}\ and\ \citenamefont
  {Logan}(2009)}]{Davidson:2009ha}%
  \BibitemOpen
  \bibfield  {author} {\bibinfo {author} {\bibfnamefont {S.~M.}\ \bibnamefont
  {Davidson}}\ and\ \bibinfo {author} {\bibfnamefont {H.~E.}\ \bibnamefont
  {Logan}},\ }\href {\doibase 10.1103/PhysRevD.80.095008} {\bibfield  {journal}
  {\bibinfo  {journal} {Phys. Rev. D}\ }\textbf {\bibinfo {volume} {80}},\
  \bibinfo {pages} {095008} (\bibinfo {year} {2009})},\ \Eprint
  {http://arxiv.org/abs/0906.3335} {arXiv:0906.3335 [hep-ph]} \BibitemShut
  {NoStop}%
\bibitem [{\citenamefont {Morrissey}\ and\ \citenamefont
  {Ramsey-Musolf}(2012)}]{Morrissey:2012db}%
  \BibitemOpen
  \bibfield  {author} {\bibinfo {author} {\bibfnamefont {D.~E.}\ \bibnamefont
  {Morrissey}}\ and\ \bibinfo {author} {\bibfnamefont {M.~J.}\ \bibnamefont
  {Ramsey-Musolf}},\ }\href {\doibase 10.1088/1367-2630/14/12/125003}
  {\bibfield  {journal} {\bibinfo  {journal} {New J. Phys.}\ }\textbf {\bibinfo
  {volume} {14}},\ \bibinfo {pages} {125003} (\bibinfo {year} {2012})},\
  \Eprint {http://arxiv.org/abs/1206.2942} {arXiv:1206.2942 [hep-ph]}
  \BibitemShut {NoStop}%
\bibitem [{\citenamefont {Weldon}(1982)}]{Weldon:1982bn}%
  \BibitemOpen
  \bibfield  {author} {\bibinfo {author} {\bibfnamefont {H.}~\bibnamefont
  {Weldon}},\ }\href {\doibase 10.1103/PhysRevD.26.2789} {\bibfield  {journal}
  {\bibinfo  {journal} {Phys. Rev. D}\ }\textbf {\bibinfo {volume} {26}},\
  \bibinfo {pages} {2789} (\bibinfo {year} {1982})}\BibitemShut {NoStop}%
\bibitem [{\citenamefont {Kobes}(1991)}]{Kobes:1990ua}%
  \BibitemOpen
  \bibfield  {author} {\bibinfo {author} {\bibfnamefont {R.}~\bibnamefont
  {Kobes}},\ }\href {\doibase 10.1103/PhysRevD.43.1269} {\bibfield  {journal}
  {\bibinfo  {journal} {Phys. Rev. D}\ }\textbf {\bibinfo {volume} {43}},\
  \bibinfo {pages} {1269} (\bibinfo {year} {1991})}\BibitemShut {NoStop}%
\bibitem [{\citenamefont {Garny}\ \emph {et~al.}(2010)\citenamefont {Garny},
  \citenamefont {Hohenegger},\ and\ \citenamefont {Kartavtsev}}]{Garny:2010nj}%
  \BibitemOpen
  \bibfield  {author} {\bibinfo {author} {\bibfnamefont {M.}~\bibnamefont
  {Garny}}, \bibinfo {author} {\bibfnamefont {A.}~\bibnamefont {Hohenegger}}, \
  and\ \bibinfo {author} {\bibfnamefont {A.}~\bibnamefont {Kartavtsev}},\
  }\href {\doibase 10.1103/PhysRevD.81.085028} {\bibfield  {journal} {\bibinfo
  {journal} {Phys. Rev. D}\ }\textbf {\bibinfo {volume} {81}},\ \bibinfo
  {pages} {085028} (\bibinfo {year} {2010})},\ \Eprint
  {http://arxiv.org/abs/1002.0331} {arXiv:1002.0331 [hep-ph]} \BibitemShut
  {NoStop}%
\bibitem [{\citenamefont {Hambye}(2002)}]{Hambye:2001eu}%
  \BibitemOpen
  \bibfield  {author} {\bibinfo {author} {\bibfnamefont {T.}~\bibnamefont
  {Hambye}},\ }\href {\doibase 10.1016/S0550-3213(02)00293-6} {\bibfield
  {journal} {\bibinfo  {journal} {Nucl. Phys. B}\ }\textbf {\bibinfo {volume}
  {633}},\ \bibinfo {pages} {171} (\bibinfo {year} {2002})},\ \Eprint
  {http://arxiv.org/abs/hep-ph/0111089} {arXiv:hep-ph/0111089} \BibitemShut
  {NoStop}%
\bibitem [{\citenamefont {Harvey}\ and\ \citenamefont
  {Turner}(1990)}]{Harvey:1990qw}%
  \BibitemOpen
  \bibfield  {author} {\bibinfo {author} {\bibfnamefont {J.~A.}\ \bibnamefont
  {Harvey}}\ and\ \bibinfo {author} {\bibfnamefont {M.~S.}\ \bibnamefont
  {Turner}},\ }\href {\doibase 10.1103/PhysRevD.42.3344} {\bibfield  {journal}
  {\bibinfo  {journal} {Phys. Rev. D}\ }\textbf {\bibinfo {volume} {42}},\
  \bibinfo {pages} {3344} (\bibinfo {year} {1990})}\BibitemShut {NoStop}%
\bibitem [{\citenamefont {Cline}\ \emph {et~al.}(1996)\citenamefont {Cline},
  \citenamefont {Kainulainen},\ and\ \citenamefont {Vischer}}]{Cline:1995dg}%
  \BibitemOpen
  \bibfield  {author} {\bibinfo {author} {\bibfnamefont {J.~M.}\ \bibnamefont
  {Cline}}, \bibinfo {author} {\bibfnamefont {K.}~\bibnamefont {Kainulainen}},
  \ and\ \bibinfo {author} {\bibfnamefont {A.~P.}\ \bibnamefont {Vischer}},\
  }\href {\doibase 10.1103/PhysRevD.54.2451} {\bibfield  {journal} {\bibinfo
  {journal} {Phys. Rev. D}\ }\textbf {\bibinfo {volume} {54}},\ \bibinfo
  {pages} {2451} (\bibinfo {year} {1996})},\ \Eprint
  {http://arxiv.org/abs/hep-ph/9506284} {arXiv:hep-ph/9506284} \BibitemShut
  {NoStop}%
\bibitem [{\citenamefont {Zyla}\ \emph {et~al.}(2020)\citenamefont {Zyla} \emph
  {et~al.}}]{Zyla:2020zbs}%
  \BibitemOpen
  \bibfield  {author} {\bibinfo {author} {\bibfnamefont {P.}~\bibnamefont
  {Zyla}} \emph {et~al.} (\bibinfo {collaboration} {Particle Data Group}),\
  }\href {\doibase 10.1093/ptep/ptaa104} {\bibfield  {journal} {\bibinfo
  {journal} {PTEP}\ }\textbf {\bibinfo {volume} {2020}},\ \bibinfo {pages}
  {083C01} (\bibinfo {year} {2020})}\BibitemShut {NoStop}%
\bibitem [{\citenamefont {Aghanim}\ \emph {et~al.}(2020)\citenamefont {Aghanim}
  \emph {et~al.}}]{Aghanim:2018eyx}%
  \BibitemOpen
  \bibfield  {author} {\bibinfo {author} {\bibfnamefont {N.}~\bibnamefont
  {Aghanim}} \emph {et~al.} (\bibinfo {collaboration} {Planck}),\ }\href
  {\doibase 10.1051/0004-6361/201833910} {\bibfield  {journal} {\bibinfo
  {journal} {Astron. Astrophys.}\ }\textbf {\bibinfo {volume} {641}},\ \bibinfo
  {pages} {A6} (\bibinfo {year} {2020})},\ \Eprint
  {http://arxiv.org/abs/1807.06209} {arXiv:1807.06209 [astro-ph.CO]}
  \BibitemShut {NoStop}%
\bibitem [{\citenamefont {Vagnozzi}\ \emph {et~al.}(2017)\citenamefont
  {Vagnozzi}, \citenamefont {Giusarma}, \citenamefont {Mena}, \citenamefont
  {Freese}, \citenamefont {Gerbino}, \citenamefont {Ho},\ and\ \citenamefont
  {Lattanzi}}]{Vagnozzi:2017ovm}%
  \BibitemOpen
  \bibfield  {author} {\bibinfo {author} {\bibfnamefont {S.}~\bibnamefont
  {Vagnozzi}}, \bibinfo {author} {\bibfnamefont {E.}~\bibnamefont {Giusarma}},
  \bibinfo {author} {\bibfnamefont {O.}~\bibnamefont {Mena}}, \bibinfo {author}
  {\bibfnamefont {K.}~\bibnamefont {Freese}}, \bibinfo {author} {\bibfnamefont
  {M.}~\bibnamefont {Gerbino}}, \bibinfo {author} {\bibfnamefont
  {S.}~\bibnamefont {Ho}}, \ and\ \bibinfo {author} {\bibfnamefont
  {M.}~\bibnamefont {Lattanzi}},\ }\href {\doibase 10.1103/PhysRevD.96.123503}
  {\bibfield  {journal} {\bibinfo  {journal} {Phys. Rev. D}\ }\textbf {\bibinfo
  {volume} {96}},\ \bibinfo {pages} {123503} (\bibinfo {year} {2017})},\
  \Eprint {http://arxiv.org/abs/1701.08172} {arXiv:1701.08172 [astro-ph.CO]}
  \BibitemShut {NoStop}%
\bibitem [{\citenamefont {Acero}\ \emph {et~al.}(2019)\citenamefont {Acero}
  \emph {et~al.}}]{Acero:2019ksn}%
  \BibitemOpen
  \bibfield  {author} {\bibinfo {author} {\bibfnamefont {M.}~\bibnamefont
  {Acero}} \emph {et~al.} (\bibinfo {collaboration} {NOvA}),\ }\href {\doibase
  10.1103/PhysRevLett.123.151803} {\bibfield  {journal} {\bibinfo  {journal}
  {Phys. Rev. Lett.}\ }\textbf {\bibinfo {volume} {123}},\ \bibinfo {pages}
  {151803} (\bibinfo {year} {2019})},\ \Eprint
  {http://arxiv.org/abs/1906.04907} {arXiv:1906.04907 [hep-ex]} \BibitemShut
  {NoStop}%
\bibitem [{\citenamefont {Machado}\ \emph {et~al.}(2015)\citenamefont
  {Machado}, \citenamefont {Perez}, \citenamefont {Sumensari}, \citenamefont
  {Tabrizi},\ and\ \citenamefont {Funchal}}]{Machado:2015sha}%
  \BibitemOpen
  \bibfield  {author} {\bibinfo {author} {\bibfnamefont {P.~A.~N.}\
  \bibnamefont {Machado}}, \bibinfo {author} {\bibfnamefont {Y.~F.}\
  \bibnamefont {Perez}}, \bibinfo {author} {\bibfnamefont {O.}~\bibnamefont
  {Sumensari}}, \bibinfo {author} {\bibfnamefont {Z.}~\bibnamefont {Tabrizi}},
  \ and\ \bibinfo {author} {\bibfnamefont {R.~Z.}\ \bibnamefont {Funchal}},\
  }\href {\doibase 10.1007/JHEP12(2015)160} {\bibfield  {journal} {\bibinfo
  {journal} {JHEP}\ }\textbf {\bibinfo {volume} {12}},\ \bibinfo {pages} {160}
  (\bibinfo {year} {2015})},\ \Eprint {http://arxiv.org/abs/1507.07550}
  {arXiv:1507.07550 [hep-ph]} \BibitemShut {NoStop}%
\bibitem [{\citenamefont {Bertuzzo}\ \emph {et~al.}(2016)\citenamefont
  {Bertuzzo}, \citenamefont {Perez~G.}, \citenamefont {Sumensari},\ and\
  \citenamefont {Zukanovich~Funchal}}]{Bertuzzo:2015ada}%
  \BibitemOpen
  \bibfield  {author} {\bibinfo {author} {\bibfnamefont {E.}~\bibnamefont
  {Bertuzzo}}, \bibinfo {author} {\bibfnamefont {Y.~F.}\ \bibnamefont
  {Perez~G.}}, \bibinfo {author} {\bibfnamefont {O.}~\bibnamefont {Sumensari}},
  \ and\ \bibinfo {author} {\bibfnamefont {R.}~\bibnamefont
  {Zukanovich~Funchal}},\ }\href {\doibase 10.1007/JHEP01(2016)018} {\bibfield
  {journal} {\bibinfo  {journal} {JHEP}\ }\textbf {\bibinfo {volume} {01}},\
  \bibinfo {pages} {018} (\bibinfo {year} {2016})},\ \Eprint
  {http://arxiv.org/abs/1510.04284} {arXiv:1510.04284 [hep-ph]} \BibitemShut
  {NoStop}%
\bibitem [{\citenamefont {Li}\ \emph {et~al.}()\citenamefont {Li},
  \citenamefont {Li}, \citenamefont {Yan},\ and\ \citenamefont
  {Yang}}]{Li-2022}%
  \BibitemOpen
  \bibfield  {author} {\bibinfo {author} {\bibfnamefont {S.-P.}\ \bibnamefont
  {Li}}, \bibinfo {author} {\bibfnamefont {X.-Q.}\ \bibnamefont {Li}}, \bibinfo
  {author} {\bibfnamefont {X.-S.}\ \bibnamefont {Yan}}, \ and\ \bibinfo
  {author} {\bibfnamefont {Y.-D.}\ \bibnamefont {Yang}},\ }\href@noop {}
  {\enquote {\bibinfo {title} {{{Effective neutrino number signals from
  keV-vacuum neutrinophilic 2HDM}}},}\ }\bibinfo {note} {{in
  preparation}}\BibitemShut {NoStop}%
\end{thebibliography}%

\end{document}